\documentclass[twocolumn,showpacs,preprintnumbers,amsmath,amssymb,numerical,prl]{revtex4-1}

\usepackage{graphicx}
\usepackage{dcolumn}
\usepackage{bm}
\usepackage{textcomp}
\usepackage{amsfonts}

\begin{document}

\title{Diodes with Breakdown Voltages Enhanced by the Metal-Insulator Transition of LaAlO$_3$-SrTiO$_3$ Interfaces}

\author{R.~Jany}
\author{M.~Breitschaft}
\author{G.~Hammerl}
\author{A.~Horsche}
\author{C.~Richter}
\author{S.~Paetel}
\author{J.~Mannhart}
\affiliation{Center for Electronic Correlations and Magnetism, Institute of Physics, University of Augsburg, 86135 Augsburg, Germany}
\author{N.~Stucki}
\author{N.~Reyren}
\author{S.~Gariglio}
\author{P.~Zubko}
\author{A.\,D.~Caviglia}
\author{J.-M.~Triscone}
\affiliation{DPMC, University of Geneva, 24 Quai E.-Ansermet, 1211 Geneva 4, Switzerland}

\date{\today}

\begin{abstract}

Using the metal-insulator transition that takes place as a function of carrier density at the LaAlO$_3$-SrTiO$_3$ interface, oxide diodes have been fabricated with room-temperature breakdown voltages of up to 200\,V. With applied voltage, the capacitance of the diodes changes by a factor of 150. The diodes are robust and operate at temperatures up to 270\,C.

\end{abstract}
                           
\maketitle

As discovered by A. Ohtomo and H. Y. Hwang, the interface between the TiO$_2$-terminated (001) surface of SrTiO$_3$ and LaAlO$_3$ can generate a conducting electron system \cite{ohtomo2004}. This system is an electron liquid \cite{breitschaft2009}, characterised by a carrier density of several $10^{13}$\,cm$^{-2}$ and a room temperature sheet resistance in the range of $10^4\,\Omega / \square$. For LaAlO$_3$ layers that are 3 or 4 unit cells (uc) thick, the electron system is close to a metal-insulator transition \cite{thiel2006, caviglia2008}. In LaAlO$_3$/SrTiO$_3$ heterostructures with a LaAlO$_3$ layer of at most 3 uc, the as-grown interface is insulating. However, an interface to a 3 uc thick LaAlO$_3$ layer becomes conducting if its carrier density is enhanced by applying a sufficiently large electric field in a field-effect transistor configuration. Correspondingly, the conductivity of an interface in a heterostructure with a 4 uc thick LaAlO$_3$ layer is switched off if the interface is depleted by a gate field.

The possibility to drive the interface through a metal-insulator transition by applying modest gate voltages is of interest for device applications.
First, the abrupt change of the interface properties at the phase transition provides the possibility to operate devices with small input voltage swings.
Second, controlling the presence of a metallic layer by switching an interface between the metallic and insulating ground state allows to effectively reconfigure a device.
Here, we report on diodes in which the latter effect is utilised to raise breakdown fields and achieve large voltage-induced capacitance changes. 

\begin{figure}[htbp]
\centering
\includegraphics[width=0.4\textwidth]{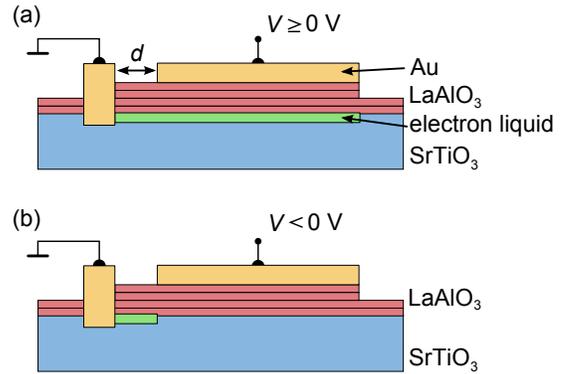}
\caption{Illustration of the operation of a device consisting of 4\,uc LaAlO$_3$ layer deposited on SrTiO$_3$.
Gold layers provide contacts to the interface and to the top of the LaAlO$_3$.
In forward direction, with a positive voltage applied to the top contact (a), a conducting electron system is formed at the LaAlO$_3$-SrTiO$_3$ interface, which is electrically separated from the top electrode by 4\,uc of LaAlO$_3$ only.
In reverse direction, with a negative voltage applied to the top electrode (b), the electron system is depleted and, undergoing a metal-insulator transition, becomes completely insulating. The effective length of the insulating region is enlarged.}
\label{img:diodenprinzip}
\end{figure}

The principle of these devices is sketched in Fig.\,\ref{img:diodenprinzip}, which shows a cross sectional cut through a self-conducting diode based on a SrTiO$_3$ ((001), TiO$_2$-terminated) - LaAlO$_3$ (4\,uc) - Au heterostructure.
One contact to the diode is provided by a gold layer on top of the LaAlO$_3$, the other contact is given by a Au-plug that fills an ion-milled hole at the side of the device.
This plug contacts the interface.
In the devices fabricated the two contacts are spaced by a lateral distance of $d \approx 1-30$\,\textmu m.
If operated with a positive voltage applied to the top contact (Fig.\,\ref{img:diodenprinzip}a), the devices resemble Schottky diodes biased in forward direction, with a current flowing by tunnelling or thermal activation across the 4\,uc thick ($\approx 1.6$\,nm) LaAlO$_3$ layer.
With a negative voltage at the top contact (Fig.\,\ref{img:diodenprinzip}b), the devices are biased in reverse direction and, for voltages exceeding $\approx 1$\,V, the metal-insulator transition is induced.
At the transition the lower electrode of the diode therefore disappears.
Thus, by applying such a voltage the effective length of the insulator in the device is enhanced from 1.6~nm to $\approx 1-30$\,\textmu m.
The diodes are therefore expected to sustain high reverse breakdown fields while supporting large forward currents.

\begin{figure}[htbp]
\centering
\includegraphics[width=0.4\textwidth]{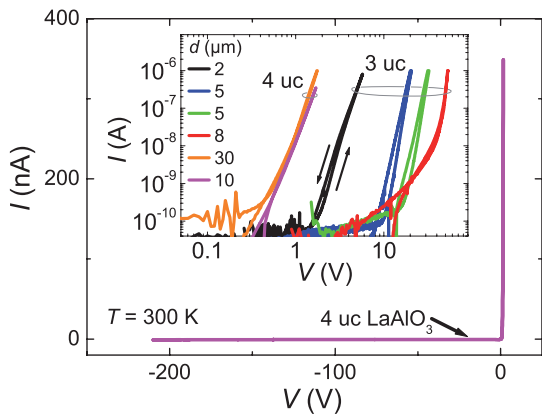}
\caption{Current-voltage characteristics of diodes as measured at $T=300$~K.
The characteristics show a rectifying behaviour with reverse breakdown voltages up to $|V|>$\,200\,V.
In the device with the 4\,uc thick LaAlO$_3$ layer, the interface is conducting at $V=0$\,V, and the device therefore has a much smaller turn-on voltage than the 3\,uc samples (inset).
The two measurements with $d=5\,$\textmu m were performed on different samples.}
\label{img:daten}
\end{figure}

Diodes with 3\,uc LaAlO$_3$ layers operate according to the same principle, the main differences being that a finite voltage in forward direction is required to switch the conducting interface on, and that due to the thinner LaAlO$_3$ larger forward current densities can be obtained.

To explore these ideas, we fabricated and experimentally investigated diodes of both types.
The devices use epitaxial LaAlO$_3$ films grown by pulsed laser deposition with reflective high energy electron diffraction as described in \cite{thiel2006}.
The LaAlO$_3$ layers were deposited at 780\,C in an oxygen pressure of 7\,$\times$\,10$^{-5}$\,mbar on TiO$_2$-terminated SrTiO$_3$ substrates \cite{kawasaki1994, koster1998} with a subsequent cooldown to 300\,K in 0.5\,bar of oxygen \cite{thiel2006}.
The devices were patterned into the geometry shown in Fig.\,\ref{img:diodenprinzip} by using the technique described in \cite{schneider2006}.
The gold top contacts were grown \textit{ex situ} using sputter deposition, then patterned by lift-off.
The contacts to the interface were provided by Ar-ion milling photolithographically defined holes which were backfilled with sputtered gold.
To avoid photoconductivity, the measurements were done after keeping the samples in dark for 24 hours. 

The current-voltage ($I$($V$)) characteristics of the devices are shown in Fig.\,\ref{img:daten}. The characteristics are stable over weeks.
As expected, the $I$($V$) curves are highly asymmetric.
In forward direction the devices have conductivities of $\approx 10^{-3}\,(\Omega\,\mathrm{cm}^{2})^{-1}$ as related to the top contact size.
The $I$($V$) characteristics show a hysteresis which we attribute to filling of trap states.
In reverse direction the diodes show breakdown voltages of several 10\,V.
In some cases the reverse breakdown voltages exceed the measurement limit of 200\,V at 300\,K.
As expected, devices with shorter $d$ have smaller switch-on voltages (forward direction).

The size of the breakdown voltage provides clear evidence that the operation of the devices relies on the field-induced metal-insulator transition.
If this was not the case the $<$\,2\,nm thick LaAlO$_3$ films would not sustain voltages $>$\,200\,V. The corresponding hypothetical electric fields of order $10^{11}$\,V/m exceed the breakdown field strength of any insulator by orders of magnitude.

\begin{figure}[htbp]
\centering
\includegraphics[width=0.4\textwidth]{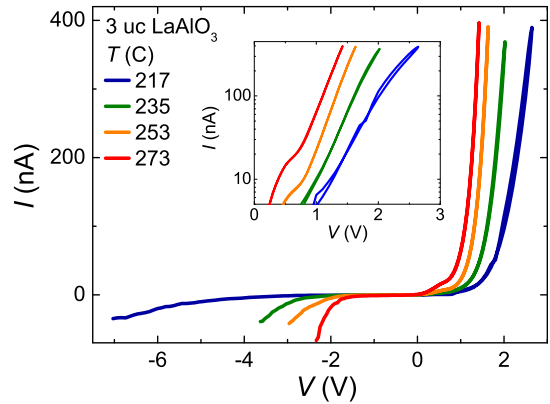}
\caption{Current-voltage characteristics of a device ($d=35\,$\textmu m) with 3\,uc of LaAlO$_3$ measured at high temperatures. }
\label{img:grafik3}
\end{figure}

To study the temperature dependence of the $I$($V$) characteristics, the devices were heated to $\approx$\,270\,C, the temperature being limited by the diffusion of the gold contacts.
To not reduce the SrTiO$_3$ the heating was done in $0.5$\,bar of O$_2$.
As shown by Fig.\,\ref{img:grafik3}, the diodes are stable and are rectifying up to the highest temperature applied.
With increasing temperature the forward current sets in at reduced voltages and the breakdown voltage in reverse direction is lowered, reaching $\approx$\,2\,V at 273\,C , in comparison to breakdown voltages of several 10\,V at room temperature.
The enhanced conduction is attributed to thermally excited charge carriers and thermally activated hopping.
It is pointed out that at high temperatures the $I$($V$) characteristics do not broaden significantly. In which manner this behavior is caused by the interface electron system, the LaAlO$_3$ barrier, or the contacts remains to be analyzed. 

\begin{figure}[htbp]
\centering
\includegraphics[width=0.4\textwidth]{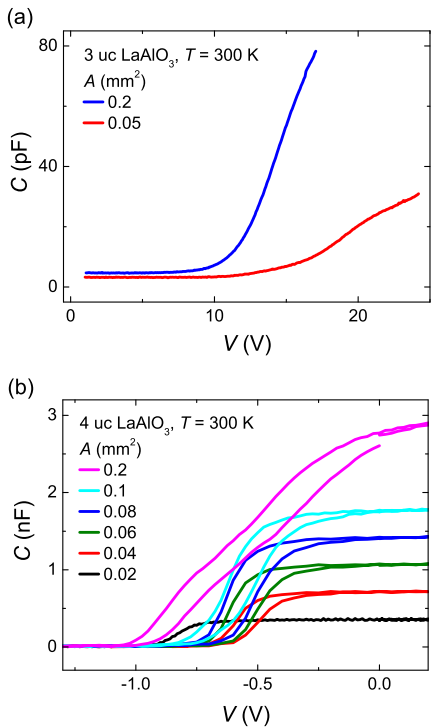}
\caption{Capacitances of devices ($d=10\,$\textmu m) with 3 and 4 uc of LaAlO$_3$ measured at 300\,K as a function of applied voltage.}
\label{img:capacity}
\end{figure}

Because the lower electrode of the device is essentially removed when the devices are switched with the applied voltage $V$ through the metal-insulator transition, the devices are expected to display a large change of capacitance $C$ with $V$.
Fig.\,\ref{img:capacity} shows the results of corresponding $C$($V$) measurements.
The capacitances of devices with 3 and 4 uc of LaAlO$_3$ were measured as a function of applied dc-voltage, using a small ac-probe voltage (40\,mV, 111\,Hz, HP 4284A {\em LCR}-meter).
As expected, the capacitance of the 4 uc diodes (Fig.\,\ref{img:capacity}b) is reduced strongly at bias voltages between 0\,V and $-1$\,V, from $\approx$\,1.35\,\textmu F/cm$^2$ to $\approx$\,0.009\,\textmu F/cm$^2$. 
Also as expected, the capacity of the self-conducting diodes scales with the area of the top contact (Fig.\,\ref{img:capacity}b).
Correspondingly, the small capacity of the unbiased, self-insulating diodes (Fig.\,\ref{img:capacity}a) is increased if the diodes are biased in forward direction.
Within the accessible voltage range, however, these diodes do not reach the capacities of the unbiased 4 uc diodes.
Similar to the $I$($V$) characteristics, also the $C$($V$) characteristics show a hysteresis, again being presumably caused by filling of trap states.
It is pointed out that in most devices the capacitance is smaller than the textbook value 

\begin{equation}
C = \varepsilon_0 \varepsilon_r \frac{A}{z},
\label{cap}
\end{equation}

where $\varepsilon_0$ and $\varepsilon_r$ are the dielectric constant of vacuum and bulk LaAlO$_3$, respectively, $A$ is the area of the top contact and $z$ the thickness of the LaAlO$_3$ film.
While most of the difference between the measured value and the one expected from Eq.\,1 is likely caused by interface states which are in particular present at the Au-LaAlO$_3$ contact, it is pointed out that we do not presume Eq.\,1 to accurately describe the capacity as besides the geometrical capacity further contributions are expected to add to the total capacity \cite{kopp2009} (see also \cite{stengel2006}). 

In summary, taking advantage of the metal-insulator transition of LaAlO$_3$-SrTiO$_3$ interfaces, oxide diodes with 
large breakdown-voltages and capacities that are highly voltage-tunable have been fabricated. The devices are robust and can be operated far above room temperature. 

The authors gratefully acknowledge helpful discussions with T.~Kopp, Y.~C.~Liao, and M.~Gleyzes. This work was supported by the DFG (TRR~80), EC (OxIDes), and by the Swiss National Sciences Foundation through the NCCR MaNEP and Division II.

\bibliographystyle{aipnum4-1}

\end{document}